\documentclass[11pt]{book}

\usepackage[dvips]{epsfig}
\usepackage{amssymb}
\usepackage{multirow}
\usepackage{amsbsy}
\usepackage{amsmath}
\usepackage{cleveref}
\usepackage{pstricks}

\catcode`\@=11

 \def\@normalsize{\@setsize\normalsize{13pt}\xipt\@xipt
   \abovedisplayskip 11pt plus3pt minus6pt
   \belowdisplayskip \abovedisplayskip
   \abovedisplayshortskip \z@ plus3pt
   \belowdisplayshortskip 6.6pt plus3.5pt minus3pt}

 \def\small{\@setsize\small{12pt}\xipt\@xipt
   \abovedisplayskip 10pt plus2pt minus5pt
   \belowdisplayskip \abovedisplayskip
   \abovedisplayshortskip \z@ plus3pt
   \belowdisplayshortskip 6pt plus3pt minus3pt
   \def\@listi{\topsep 6pt plus 2pt minus 2pt
     \parsep 3pt plus 2pt minus 1pt
     \itemsep \parsep}}

 \def\footnotesize{\@setsize\footnotesize{10pt}\ixpt\@ixpt
   \abovedisplayskip 8pt plus 2pt minus 4pt
   \belowdisplayskip \abovedisplayskip
   \abovedisplayshortskip \z@ plus 1pt
   \belowdisplayshortskip 4pt plus 2pt minus 2pt
   \def\@listi{\topsep 4pt plus 2pt minus 2pt
      \parsep 2pt plus 1pt minus 1pt
      \itemsep \parsep}}

 \def\scriptsize{\@setsize\scriptsize{9.5pt}\viiipt\@viiipt}
 \def\tiny{\@setsize\tiny{7pt}\vipt\@vipt}
 \def\large{\@setsize\large{14pt}\xiipt\@xiipt}
 \def\Large{\@setsize\Large{18pt}\xivpt\@xivpt}
 \def\LARGE{\@setsize\LARGE{22pt}\xviipt\@xviipt}
 \def\huge{\@setsize\huge{25pt}\xxpt\@xxpt}
 \def\Huge{\@setsize\Huge{30pt}\xxvpt\@xxvpt}

\def\section{\@startsection {section}{1}{\z@}%
{-1.5\baselineskip plus-1pt minus-3pt}{1\baselineskip plus1pt minus2pt}%
{\centering\normalsize\bf}}
\def\subsection{\@startsection{subsection}{2}{\z@}%
{-1\baselineskip plus-1pt minus-2pt}{1\baselineskip plus1pt minus2pt}%
{\normalsize\sc\noindent}}
\def\subsubsection{\@startsection{subsubsection}{3}{\z@}%
{-1\baselineskip plus-1pt minus-2pt}{1sp}{\normalsize\it\noindent}}
\def\paragraph{\@startsection{paragraph}{4}{\z@}%
{1\baselineskip plus1pt minus2pt}{-1em}{\normalsize\it\noindent}}
\let\subparagraph=\paragraph

\def\tableofcontents{\@restonecolfalse\if@twocolumn\@restonecoltrue
\onecolumn\fi\OSIDcont\@starttoc{con}\if@restonecol\twocolumn\fi}

\def\l@section{\@dottedtocline{1}{0em}{.66em}}

\def\thebibliography#1{\section*{{Bibliography}\@mkboth
 {BIBLIOGRAPHY}{BIBLIOGRAPHY}}\footnotesize\rm\list
 {[\arabic{enumi}]}{\settowidth\labelwidth{[#1]}\leftmargin\labelwidth
 \advance\leftmargin\labelsep\usecounter{enumi}}
 \def\newblock{\hskip .11em plus .33em minus -.07em}
 \sloppy\clubpenalty4000\widowpenalty4000
 \sfcode`\.=1000\relax}

\def\ps@myheadings{\let\@mkboth\@gobbletwo
\def\@oddhead{\hfil{\footnotesize\rm\rightmark}\hfil}
\def\@evenhead{\hfil{\footnotesize\rm\leftmark}\hfil}
\def\@oddfoot{\hfil{\footnotesize\sf\artid-\thepage}\hfil}
\def\@evenfoot{\hfil{\footnotesize\sf\artid-\thepage}\hfil}
\def\sectionmark##1{}\def\subsectionmark##1{}}

\def\@copyrighthead{\parbox{127mm}{\footnotesize\rm\ \\[6pt]
Open Systems~\& Information Dynamics\\
Vol.~\Vol, No.~\Number~(\Year)~\artid~(\EndpagE~pages)\\
DOI:\DOInumber\\
\copyright~World Scientific Publishing Company\\
\epsfxsize=4cm
\vskip-\lastskip
\vskip-\baselineskip
\vspace*{-38.5pt}
}}

\def\artid{0000001}
\def\Year{2008}        %
\def\Vol{15}           
\newcounter{paPer}     %
\def\EndpagE{\expandafter\pageref{\the\value{paPer}OpSy}}

\def\ps@osiD{\let\@mkboth\@gobbletwo
\def\@oddhead{\@copyrighthead}
  \def\@oddfoot{\hfil{\footnotesize\sf\artid-\thepage}\hfil}
  \def\@evenhead{}\let\@evenfoot\@oddfoot}

\def\cite{\@ifnextchar [{\@tempswatrue\@Rcitex}{\@tempswafalse\@Rcitex[]}}

\def\@Rcitex[#1]#2{\if@filesw\immediate\write\@auxout{\string\citation{#2}}\fi
  \def\@citea{}\@cite{\@for\@citeb:=#2\do
    {\@citea\def\@citea{,\penalty\@m\,}\@ifundefined
       {b@\the\value{paPer}R\@citeb}{{\bf ?}\@warning
       {Citation `\@citeb' on page \thepage \space undefined}}%
\hbox{\csname b@\the\value{paPer}R\@citeb\endcsname}}}{#1}}

\long\def\@caption#1[#2]#3{\par\addcontentsline{\csname
  ext@#1\endcsname}{#1}{\protect\numberline{\csname
  the#1\endcsname}{\ignorespaces #2}}\begingroup
    \@parboxrestore
    \small                                        
    \@makecaption{\csname fnum@#1\endcsname}{\ignorespaces #3}\par
  \endgroup}

\newtoks\@stequation

\def\subequations{\refstepcounter{equation}%
\edef\@savedequation{\the\c@equation}%
\@stequation=\expandafter{\theequation}
\edef\@savedtheequation{\the\@stequation}
\edef\oldtheequation{\theequation}%
\setcounter{equation}{0}%
\def\theequation{\oldtheequation\alph{equation}}}%

\def\endsubequations{%
\setcounter{equation}{\@savedequation}%
\@stequation=\expandafter{\@savedtheequation}%
\edef\theequation{\the\@stequation}\global\@ignoretrue}

\catcode`\@=12

\let\Rlabel=\label
\let\Rbibitem=\bibitem
\let\Rref=\ref
\let\Rpageref=\pageref
\def\label#1{\expandafter\Rlabel{\the\value{paPer}R#1}}
\def\bibitem#1{\expandafter\Rbibitem{\the\value{paPer}R#1}}
\def\ref#1{\expandafter\Rref{\the\value{paPer}R#1}}
\def\pageref#1{\expandafter\Rpageref{\the\value{paPer}R#1}}

\def\thesection{\arabic{section}.}

\def\YYMm{\rule{0ex}{4em}}
\newtoks\TITsi
\newtoks\TITsii

\def\title#1{\def\TITs{\LARGE{\raggedright\noindent\YYMm #1%
\vskip8pt\par}}}

\def\author#1{\autMM{#1}\def\LHD{#1}}
\def\and{{\rm\lowercase{and}}}

\def\autMM#1{\TITsii={\vskip10pt\par\normalsize\rm\noindent #1\par}%
\TITsi=\expandafter{\TITs}\edef\TITs{\the\TITsi\the\TITsii}}

\def\address#1{\TITsii={\vskip6pt\par\footnotesize\sl\noindent #1\par}%
\TITsi=\expandafter{\TITs}%
\edef\TITs{\the\TITsi\the\TITsii}}

\def\received#1{\TITsii={\vskip10pt\par\small\rm\noindent(Received: #1)\par}%
\TITsi=\expandafter{\TITs}\edef\TITs{\the\TITsi\the\TITsii}}

\def\headtitle#1{\def\RHD{#1}}
\def\headauthor#1{\def\LHD{#1}}
\def\listas#1#2{\addcontentsline{con}{section}{{\sc #1: }{\rm #2}}}

\def\abst{{\bf Abstract.}}
\def\abstract#1{\TITs
       \vskip15pt\par\noindent
       {\footnotesize{\abst~} #1\vskip3pt\par}
       \markright{\RHD}
       \markboth{\LHD}{\RHD}}

\def\startpaper{%
       \cleardoublepage
       \setcounter{section}{0}
       \stepcounter{paPer}
       \setcounter{equation}{0}
       \setcounter{footnote}{0}
       \setcounter{figure}{0}
       \setcounter{table}{0}
       \def\theequation{\arabic{equation}}
       \def\thefootnote{\arabic{footnote}}
       \setcounter{defn}{0}
       \setcounter{thm}{0}
       \setcounter{lem}{0}
       \setcounter{prop}{0}
       \setcounter{rem}{0}
       \thispagestyle{osiD}}

\def\OSIDcont{\cleardoublepage\thispagestyle{empty}
       \markright{}\markboth{}{}
       \normalsize\rm
       \hspace*{\fill}{\large\rm
         Contents of the Volume \Volume, Number \Number}\hspace*{\fill}
       \par\vspace{1.5em}
       \par\noindent}

\def\endpaper{\expandafter\label{\the\value{paPer}OpSy}}

\def\1{{\mathchoice{\rm 1\mskip-4mu l}{\rm 1\mskip-4mu l}%
{\rm 1\mskip-4.5mu l}{\rm 1\mskip-5mu l}}}

\def\varkappa{\mbox{\bBB\char 123}}

\def\longhookrightarrow{\lhook\joinrel\relbar\joinrel\rightarrow}

\def\longhookUp{\lower6pt\hbox{\rotatebox{90}{$\longhookrightarrow$}}}

\def\theequation{\thesection\arabic{equation}}

\def\Myskip{\setlength{\baselineskip}{13pt}}

\def\text#1{\quad\mbox{\rm  #1 }\quad}

\def\DOInumber{}

\input xy
\xyoption{all}

\InputIfFileExists{psfig.sty}{\typeout{^^Jpsfig.sty inputed...ok}}{\typeout{^^JWarning: psfig.sty could not be found.^^J}}

\begin{document}

\def\artid{0000001}
\def\Volume{}
\def\Number{}
\def\Year{}
\setcounter{page}{1}

\def\DOInumber{}

\startpaper

\newcommand{\Mn}{M_n(\mathbb{C})}
\newcommand{\Mk}{M_k(\mathbb{C})}
\newcommand{\id}{\mbox{id}}
\newcommand{\ot}{{\,\otimes\,}}
\newcommand{{\Cd}}{{\mathbb{C}^d}}
\newcommand{\sbsigma}{{\mbox{\scriptsize \boldmath $\sigma$}}}
\newcommand{\sbalpha}{{\mbox{\scriptsize \boldmath $\alpha$}}}
\newcommand{\sbbeta}{{\mbox{\scriptsize \boldmath $\beta$}}}
\newcommand{\bsigma}{{\mbox{\boldmath $\sigma$}}}
\newcommand{\balpha}{{\mbox{\boldmath $\alpha$}}}
\newcommand{\bbeta}{{\mbox{\boldmath $\beta$}}}
\newcommand{\bmu}{{\mbox{\boldmath $\mu$}}}
\newcommand{\bnu}{{\mbox{\boldmath $\nu$}}}
\newcommand{\ba}{{\mbox{\boldmath $a$}}}
\newcommand{\bb}{{\mbox{\boldmath $b$}}}
\newcommand{\sba}{{\mbox{\scriptsize \boldmath $a$}}}
\newcommand{\MD}{\mathfrak{D}}
\newcommand{\de}{\Delta}
\newcommand{\sbb}{{\mbox{\scriptsize \boldmath $b$}}}
\newcommand{\sbmu}{{\mbox{\scriptsize \boldmath $\mu$}}}
\newcommand{\sbnu}{{\mbox{\scriptsize \boldmath $\nu$}}}
\def\oper{{\mathchoice{\rm 1\mskip-4mu l}{\rm 1\mskip-4mu l}%
{\rm 1\mskip-4.5mu l}{\rm 1\mskip-5mu l}}}
\def\<{\langle}
\def\>{\rangle}
\def\theequation{\thesection\arabic{equation}}

\catcode`\>=\active \def>{
\fontencoding{T1}\selectfont\symbol{62}\fontencoding{\encodingdefault}}
\catcode`\|=\active \def|{
\fontencoding{T1}\selectfont\symbol{124}\fontencoding{\encodingdefault}}
\newcommand{\mathd}{\mathrm{d}}
\newcommand{\mathe}{\mathrm{e}}
\newcommand{\nobracket}{}
\newcommand{\nocomma}{}
\newcommand{\tmfolded}[2]{\trivlist{\item[$\bullet$]\mbox{}#1}}
\newcommand{\tmfoldedplain}[2]{\trivlist{\item[]\mbox{}#1}}
\newcommand{\tmop}[1]{\ensuremath{\operatorname{#1}}}
\newcommand{\tmsamp}[1]{\textsf{#1}}
\newcommand{\tmem}[1]{{\em #1\/}}
\newcommand{\tmtextbf}[1]{{\bfseries{#1}}}

\newcommand{\ket}[1]{{\left\vert {#1} \right\rangle}}
\newcommand{\bra}[1]{{\left\langle {#1} \right\vert}}
\newcommand{\bracket}[2]{\langle{#1}|{#2}\rangle}
\newcommand{\ketbra}[1]{|{#1}\rangle\langle{#1}|}
\newcommand{\proj}[1]{|{#1}\rangle\langle{#1}|}
\newcommand{\eq}{Eq.~}
\newcommand{\eqs}{Eqs.~}
\newcommand{\fig}{Fig.~}
\newcommand{\figs}{Figs.~}
\newcommand{\cf} {cf.~}
\newcommand{\ug} {\!=\!}
\newcommand{\tens} {\!\otimes\!}
\newcommand{\piu} {\!+\!}
\newcommand{\meno} {\!-\!}
\newcommand{\ie} {i.e.~}
\newcommand{\eg} {e.g.~}
\newcommand{\av}[1]{\langle#1\rangle}
\newcommand{\up}{\uparrow}
\newcommand{\dow}{\downarrow}
\newcommand{\rref} {Ref.~}
\newcommand{\rrefs} {Refs.~}
\newcommand{\kket}[1]{\left.\left\vert#1\right\rangle\right\rangle}
\newcommand{\bbra}[1]{\left\langle\left\langle#1\right\vert\right.}
\newcommand{\U}{\mathcal{U}_\tau}
\newcommand{\J}{\mathcal{J}}
\newcommand{\R}{\mathcal{R}}
\newcommand{\sw}{\mathcal{S}_{1A}}
\newcommand{\compcent}[1]{\vcenter{\hbox{$#1\circ$}}}

\newcommand{\comp}{\mathbin{\mathchoice
  {\compcent\scriptstyle}{\compcent\scriptstyle}
  {\compcent\scriptscriptstyle}{\compcent\scriptscriptstyle}}}
\newcommand{\limnt}{\lim_{\substack{n\rightarrow\infty\\\tau\rightarrow 0}}}

\title{Stochastic versus periodic quantum collision models}

\author{Francesco Ciccarello}
\address{Universit$\grave{a}$  degli Studi di Palermo, Dipartimento di Fisica e Chimica -- Emilio Segr$\grave{e}$, via Archirafi 36, I-90123 Palermo, Italy}
\address{NEST, Istituto Nanoscienze-CNR, Piazza S. Silvestro 12, 56127 Pisa, Italy}

\headauthor{}
\headtitle{}
\received{}
\listas{autori}{titolo}

\abstract{Most literature on quantum collision models (CMs) usually considers periodic weak collisions featuring a fixed waiting time between two next collisions. Some works have yet addressed CMs with random waiting time and strong collisions (stochastic CMs). This short paper discusses how the open dynamics arising from these two types of models can be formally mapped with one another. This can be achieved for a given stochastic CM by constructing an associated periodic CM such that the waiting time randomness of the former is turned into the mixedness of the ancilla's initial state of the latter, introducing at the same time an additional state of the ancilla. Through this mapping, non-Markovian behaviour arising from a constrained number of stochatic collisions can be linked to initial ancilla-ancilla correlations of the associated periodic CM.}

\Myskip



\section{Introduction} \label{intro}
\setcounter{equation}{0}

In the last few years, quantum collision models (CMs) have gained growing popularity within the general framework of open quantum system theory \cite{campbell2021collision,ciccarello2022quantum}. At variance with traditional system-reservoir models, a CM assumes the bath decomposed into non-interacting smaller units (ancillas) which interact one at a time with the open system $S$. The reduction of the complex system-bath dynamics to a sequence of simple two-body interactions (collisions) has several advantageous applications.

The most common, and in some respects simpler, version of CM considers {\it periodic} collisions in the sense that the time between two next collisions (waiting time) is fixed. Most often, the waiting time is assumed to be zero and each collision to last a fixed time $\de t$. In this framework, the system-ancilla interaction is typically thought as weak, hence the effect of the bath on the open system is somehow diluted in time.

In stark contrast, some works in the CM literature \cite{vacchini_non-markovian_2013,vacchini_general_2014,lorenzo_quantum_2017,BrunnerNJP15,ScaraniPRE2019,ChisholmPRR2020,vacchini2020quantum,ChisholmNJP,chisholm2021witnessing,o2021stochastic} assume a stochastic waiting time, often assumed to follow a Poisson probability distribution function, and strong system-ancilla interactions. Unlike periodic CMs, in such stochastic CMs the effect of the bath is highly concentrated at the random times at which collisions occur. In particular, so long as one expects the CM to describe a gas-like bath, assuming a random (instead of fixed) waiting time appears physically motivated\footnote{On the other hand, there exist experimentally-relevant dynamics, such as some important quantum optics phenomena, which are instead naturally mapped into CMs with periodic collision time \cite{ciccarello2017collision,cilluffo2020}.}. 

While both the aforementioned types of models (respectively periodic and stochastic CMs) lead to a master equation (ME) in Lindblad form, there are interesting differences such as the physical origin of the rate and jump operators entering the ME dissipator and the quantum map describing the effect of each collision on the system. 

Recently, a variety of aspects and properties of quantum CMs were settled within a general theoretical framework, this  yet being mostly built on the paradigm of periodic CMs \cite{ciccarello2022quantum}. It is natural to ask whether and how stochastic CMs could be accommodated in such general theory or whether they instead require an {\it ad hoc} treatment. 
In this short paper, we discuss how such a formal incorporation is indeed possible and can be carried out in a rather simple and physically intuitive way. This is achieved by somehow turning the waiting time's randomness into the mixedness of the initial state of a set suitably-defined ancillas, one for each time interval. We note that a similar construction was presented in \rref\cite{gross2018qubit} in a weak continuous measurements framework. Here, this mapping is reconsidered from a more somewhat pedagogical perspective with the main goal of accommodating it in the general theory of CMs \cite{ciccarello2022quantum} also in view of possible non-Markovian generalizations (which will be discussed in the final part).

The paper starts in Section 2 with a short review of a paradigmatic CM with stochastic waiting time and the derivation of the ensuing ME. Next, Section 3 shows how to construct a corresponding CM with periodic collisions which returns the same open dynamics as the stochastic CM. Finally, Section 4 presents a summary and discussion of the results.

\section{Collision model with stochastic waiting time} \label{review1}


Consider a quantum system $S$ and a bath (\ie large collection) of identical non-interacting ancillas. By hypothesis, ancillas are all prepared in the same initial state $\eta$ and they are fully uncorrelated with one another as well as with the open system $S$.

Occasionally, a single ancilla of the bath undergoes a binary ``collision" with system  $S$. Such a collision is described by a bipartite unitary $U$ acting on the joint Hilbert space of $S$ and the colliding ancilla. If $\rho$ is the state of $S$ before the collision starts, at the end of the collision its reduced state is given by
\begin{equation}
	\rho'={\Lambda} [\rho]\label{rhop}\,.
\end{equation}
Here, quantum map ${\Lambda} $ is defined by
\begin{equation}
	{\Lambda} [\rho]={\rm Tr}_{\rm anc} \{U \rho \otimes \eta\, U^\dag \}\label{map}\,,
\end{equation}
where ${\rm Tr}_{\rm anc}$ denotes the partial trace over the ancilla's degrees of freedom. 

The waiting time $T$ between two subsequent system-ancilla collisions follows a standard exponential distribution
\begin{equation}
	p(T)= \gamma e^{-\gamma T}\,
\end{equation}
with $p(T)$ the probability of having two next collisions separated by a time $T$.
The {\it collision rate} $\gamma$ represents the characteristic rate of occurrence of collisions, its rigorous definition being
\begin{equation}\label{dp}
	\Delta p = \gamma \Delta t\,\,\,\,\,\,\,({\rm for}\,\,\,\Delta t\rightarrow 0)
\end{equation}
with $\Delta p$ the probability that an ancilla collides with $S$ in a very short time window of width $\Delta t$\footnote{For presentation issues, we avoid writing explicitly infinitesimal quantities (\eg $dp$ in place of $\Delta p$) to better highlight formal analogies with the collision model introduced in the following section.}. The fact that, irrispective of the considered time interval and past history, the collision probability is always given by \eqref{dp} should be acknowledged as a generally non-obvious condition linked to the Markovian nature of the dynamics as we discuss shortly.

To derive the master equation (ME) for $S$ we observe that, in each time interval $\Delta t$, system $S$ can either collide with an ancilla so as to change its state according to \eqref{map} or remain unchanged (since in such a case no collision has occured in the considered time window). The former event occurs with probability $ \de p$ [\cf\eq\eqref{dp}] while the latter with probability $1-\de p$. Hence, on average, the state of $S$ at the end of the short time interval is\footnote{For simplicity, we assume that $H_S=0$, \ie $S$ lacks a free Hamiltonian.} 
\begin{equation}
	\rho'=(1-\Delta p)\rho+\Delta p \,\Lambda[\rho]\,.\label{rhop}
\end{equation}

Note \eq\eqref{rhop} can be arranged compactly as the action of a quantum map as
\begin{equation}
	\rho'=\tilde{\Lambda}[\rho]\,\,\,\,{\rm with}\,\,\,\tilde{\Lambda}[\rho]=(1-\de p)\rho+\de p \,\Lambda[\rho]\,.\label{map22}
\end{equation}
Defining $\de \rho=\rho'-\rho$ and dividing either side by $\de t$, this equation can be rearranged as
\begin{equation}
	\frac{\de \rho}{\de t}=\frac{\de p}{\de t} \left( \Lambda[\rho]-\rho\right)\,.\label{ME}
\end{equation}
Finally, taking the continuous-time limit $\de t\rightarrow 0$ and introducing rate $\gamma$ through \eq\eqref{dp}, we get the ME
\begin{equation}
	\dot \rho=\gamma \left( \Lambda[\rho]-\rho\right)\,.\label{ME2}
\end{equation}
By decomposing map $\Lambda$ in terms of Kraus operators $\{K_\nu\}$, it is easily shown (see Appendix) that ME \eqref{ME} can be arranged in the standard Lindblad form
\begin{equation}
	\dot \rho=\gamma \sum_\nu \left( K_\nu \rho K_\nu^\dag -\tfrac{1}{2}\left[ K_\nu^\dag K_\nu,\rho\right]_+\right)\,,\label{ME3}
\end{equation}
where $[...,...]_+$ is the anti-commutator.

As pointed out in \rref\cite{ScaraniPRE2019}, an interesting property of the present stochastic CM is that the collision rate $\gamma$ coincides with the rate entering the dissipator of ME \eqref{ME}. This is rather different from standard periodic CMs \cite{ciccarello2022quantum}, where the dissipation rate depends instead on the interaction strength of the system-ancilla interaction Hamiltonian and on the collision duration. In contrast, in master equation \eqref{ME3} the dependence on the parameters of the system-ancilla collision is fully contained inside jump operators $\{K_\nu\}$ which in the present case coincide with the Kraus operators of the collision map $\Lambda$.

\section{Mapping into a collision model with periodic collisions} \label{map-1}

Based on the stochastic CM in the previous section, we next define an associated CM yet featuring {\it periodic} collisions (namely fixed deterministic waiting time between two next collisions). In order to simplify the language, we will conveniently refer to the CM to be defined as ``new" and the stochastic one in the previous section as  ``old".

The new CM again features a bath of identical uncoupled ancillas. The generic ancilla yet differs from that in the old CM for the inclusion of an auxiliary state. By this, we mean that if $\{\ket{m}\}$ is a basis of the old ancilla the new ancilla basis is instead given by $\{\ket{\phi},\,\{\ket{m}\}\}$, where $\ket{\phi}$ is the extra auxiliary state\footnote{Thereby, if $d$ (integer number) is the Hilbert space dimension of the old ancilla, the new ancilla is a $(d{+}1)$--dimensional system.}.

By hypothesis, a collision of the new ancilla with $S$ is described by a unitary $\tilde U$ that acts trivially on $S$ and the ancilla when the latter is in state $\ket{\phi}$, being otherwise identical to the collision unitary $U$ of the old CM. Accordingly, the new collision unitary $\tilde U$ has the form
\begin{equation}
	\tilde U=  I_S \otimes \ket{\phi}_{\rm anc}\!\langle \phi | + U\,\,.\label{U}
\end{equation}
with $U$ having support on the tensor product of the Hilbert space of $S$ and the ancilla's supsbace spanned by $\{\ket{m}\}$, which in particular entails $U|\phi\rangle=0$.

Therefore, $S$ is insensitive to the ancilla when this is in state $\ket{\phi}$.
In practice, as an instance one could think of the old ancilla as a two-level atom with ground (excited) state $\ket{g}$ ($\ket{e}$) and the new ancilla as the same atom where an additional excited level $\ket{\phi}$ is included (\eg according to a V-type configuration such that only transitions $\ket{g}\leftrightarrow \ket{e}$ and $\ket{g}\leftrightarrow \ket{\phi}$ are allowed). If transition $\ket{g}\leftrightarrow \ket{\phi}$ is far-detuned from the characteristic frequency of $S$ then only states $\{\ket{m}\}$ with $m=g,e$ will be involved in the collision unitary\footnote{In an alternative more elegant implementation, $\ket{e}$ and $\ket{\phi}$ have the same energy. If so, then there exists a dark state, which is a superposition of $\ket{e}$ and $\ket{\phi}$, that remains uncoupled from $S$.}. As is probably clear already, the auxiliary state $\ket{\phi}$ is intended to simulate the {\it non}-occurrence of a system-ancilla collision. 

Consider now a mesh of time defined by $t_n=n\Delta t$ with $n$ integer. For any time interval, each of duration $\Delta t$ (that should be thought as very short), we consider an ancilla of the new CM as the one just defined above. Ancillas are thus labeled by $n$ so that ancillas with different values of $n$ are mutually commuting.

By hypothesis of the model, each ancilla is initially in the mixed state
\begin{equation}
	\tilde \eta= (1-\Delta p)\,\ket{\phi}\langle \phi|+\Delta p\,\eta\label{chi}
\end{equation}
with $\Delta p=\gamma \de t$ just as in \eq\eqref{dp}. Here, $\eta$ is fully contained in the subspace spanned by $\{\ket{m}\}$, hence it is orthogonal to $\ket{\phi}$.
Note that, since $\Delta p$ is small, state \eqref{chi} should be viewed as a mixture dominated by state $\ket{\phi}$ plus a little component $\eta$. This entails that, as the former state is uncoupled from $S$, each single ancilla just defined almost surely will not collide with the system in the corresponding (short) time interval of duration $\Delta t$. This is in contrast to the stochastic (\ie old) CM in the previous section where instead the collision has an impulsive nature.

The collision map for this new CM is now worked out in formal analogy with \eqref{map} by replacing $\eta$ with $\tilde \eta$ and $U$ with $\tilde U$ [\cf\eqs\eqref{U} and \eqref{chi}], which gives
\begin{align}
	{\tilde \Lambda} [\rho]&={\rm Tr}_{\rm anc} \{\tilde U \rho \otimes \tilde \eta\, \tilde U^\dag \}=(1-\Delta p) \rho +\Delta p\, {\rm Tr}_{\rm anc} \{U \rho \otimes \eta\, U^\dag \}\nonumber\\
	&=(1-\Delta p) \rho +\Delta p\,\Lambda [\rho]\,.\label{map2}
\end{align}
This is just the same quantum map as \eqref{map22}, which shows that the new CM with periodic collisions reproduces exactly the same open dynamics as the old CM with stochastic collisions.

\section{Discussion}\label{disc-sec}

We have considered a paradigmatic stochastic quantum collision model (CM). Such models feature impulsive collisions separated by a random waiting time at variance with standard CMs consisting instead of weak periodic collisions. Given a stochastic CM, one can always build up an associated periodic CM yielding the same open dynamics. This is defined by introducing an additional ancillary state $\ket{\phi}$, which is insensitive to the presence of the open system $S$, and preparing each ancilla in a mixed state comprising a dominant contribution from an inhert state $\ket{\phi}$ and small one from a state $\eta$ that does interact with $S$. The rationale of this is that the occasional but impulsive collision in a stochastic CM translates into a minority of ancillas of a periodic CM which are prepared in a state stronly sensitive to the system (while most ancillas are inert). This scheme is reminiscent of the standard decomposition of the field into time bins which is a routine description in photon-counting statistics theories \cite{fox2006quantum}. 

Recently, \rref\cite{ChisholmPRR2020} showed that enforcing a {\it fixed} number of collisions in a stochastic CM generally results in a non-Markovian dynamics of $S$ as opposed to a Poissonian distribution of waiting times (corresponding to an uncertain number of collisions and, as reviewed above, Markovian behaviour). It is natural to ask whether, based on the above mapping, such non-Markovianity mechanism can be connected with one of the three basic mechanisms of introducing memory in a {\it periodic} CM discussed in \rref\cite{ciccarello2022quantum} (ancilla-ancilla collisions, initially correlated ancillas, multiple collisions). 

To this aim, we note that, clearly, in the periodic CM that we constructed the probability to have $m$ ancillas in state $\eta$ (giving rise to a collision) follows a Poissonian distribution (with $1\le m\le [t_{\rm max}/\Delta t]$ where $t_{\rm max}$ is the total evolution time). This crucially relies on the hypothesis that ancillas share no initial correlations (the bath starts in the product state $\otimes_n \tilde \eta_n$), which leads to a fully Markovian dynamics for $S$ \cite{ciccarello2022quantum}. 
If instead the number of collisions in the original stochastic CM were fixed, then in the mapped periodic CM such constraint would necessarily translate into initially correlated ancillas\footnote{For instance, if (for the sake of argument) we take 3 ancillas, an initial bath state yielding exactly two collisions could be the mixture $\rho_B=p_1\, \ket{\phi}_1\!\bra{\phi} \,\eta_2\,\eta_3+p_2\, \eta_1 \,\ket{\phi}_2\!\bra{\phi}\,\eta_3+p_3\, \eta_1\, \eta_2\,\ket{\phi}_3\!\bra{\phi}$ with $p_1+p_2+p_3=1$.}. This suggests that non-Markovianity due to a constrained number of stochastic collisions is ultimately connected to the known memory mechanism stemming from initially-correlated ancillas \cite{ciccarello2022quantum,filippov_divisibility_2017}.

\section*{Acknowledgments}
We gratefully acknowledge D.~Cilluffo, S.~Lorenzo, B.~Vacchini and D.~Chisholm for very useful discussions and the critical reading of the manuscript.

Financial support from MUR through project PRIN (Project No. 2017SRN-BRK QU- SHIP) is acknowledged.

\section*{Appendix}

Since it is completely positive and trace preserving, map $\Lambda$ can be decomposed as
\begin{equation}
	\Lambda [\rho]= \sum_\nu K_\nu \rho K_\nu^\dag
\end{equation}
with $\{K_\nu\}$ a set of Kraus operators fulfilling $\sum_\nu K_\nu^\dag K_\nu =\mathbb{I}_S$. This last equation entails
\begin{equation}
	\rho=\tfrac{1}{2}\left(\mathbb{I}_S\rho+\rho \mathbb{I}_S\right)=\sum_\nu \tfrac{1}{2 }\left[K_\nu^\dag K_\nu,\rho\right]_+\,.
\end{equation}
Replacing this in the right hand side of \eqref{ME2} we end up with \eq\eqref{ME3}.

\bibliography{torun_bib}{}
\bibliographystyle{ieeetr}

\endpaper

\end{document}